\documentclass{ws-ijmpa}
\def\ra{\rightarrow}
\def\be{\begin{equation}}
\def\ee{\end{equation}}
\def\bea{\begin{eqnarray}}
\def\eea{\end{eqnarray}}

\newcommand{\ted}{$\theta_{13}$}

\newcommand{\bnel}{\mbox{$\bar{\nu}_e$} }
\newcommand{\bnmu}{\mbox{$\bar{\nu}_\mu$} }

\newcommand{\nel}{\mbox{$\nu_e$} }
\newcommand{\nmu}{\mbox{$\nu_\mu$} }
\newcommand{\ntau}{\mbox{$\nu_\tau$} }
\newcommand{\sk}{Super-Kamiokande }
\newcommand{\neu}{neutrino }
\newcommand{\neus}{neutrinos }

\newcommand{\lsnd}{LSND }

\newcommand{\oscs}{oscillations }
\newcommand{\majo}{Majorana }
\newcommand{\ema}{\mbox{$\langle m_{\nu_e} \rangle$ }}
\newcommand{\omnu}{$\Omega_\nu h^2$ }
\newcommand{\exps}{experiments }
\newcommand{\sint}{\mbox{$\sin^2 2\theta$} }
\newcommand{\delm}{\mbox{$\Delta m^2$} }

\begin{document}

\markboth{Kai Zuber}
{Experimental Neutrino Physics}

%
\catchline{}{}{}{}{}
%

\title{EXPERIMENTAL NEUTRINO PHYSICS}

\author{K. Zuber}

\address{Denys Wilkinson Laboratory, Keble Road, University of Oxford, Oxford OX1 3RH \\
Dept. of Physics and Astronomy, University of Sussex, Falmer, Brighton BN1 9QH}

\maketitle

\pub{Received (Day Month Year)}{Revised (Day Month Year)}

\begin{abstract}
The current experimental status of neutrino physics is reviewed. It contains the 
evidences for a non-vanishing neutrino rest mass from neutrino oscillation searches.
In addition an outlook is given on determining the various mixing matrix elements and mass differences
more precisely with new experiments. Of special interest is the value of the mixing angle \ted{} determining
the possibility of detecting leptonic CP violation in the future. The prospect for absolute mass measurements
using beta and double beta decay as well as cosmological observations is presented.
\end{abstract}

\section{Introduction}	
In the last decade convincing evidence has been found for a non-vanishing rest mass
of neutrinos. 
If \neus are massive the weak and mass eigenstates are not necessarily identical, a fact
well known in the quark sector where both types of states are connected
by the CKM-matrix.
This would allow for a similar mixing matrix in the leptonic sector called PMNS-matrix and 
for the phenomenon of neutrino oscillations, a kind of flavour
oscillation, which is already known in other particle systems.

\section{Evidence for \neu \oscs and tests in the near future}
Currently we have three evidences for \neu \oscs coming from accelerators,
the atmosphere and the Sun. All evidences will be discussed in a two flavour
scenario, where the mixing is described by
\begin{equation}
{\nu_e \choose \nu_\mu} = \left( \begin{array}{cc}
 \cos \theta & \sin \theta \\
- \sin \theta & \cos \theta
\end{array} \right)
{\nu_1 \choose \nu_2}
\end{equation}
with $\theta$ as the mixing angle, analoguous to the Cabibbo angle in the quark
mixing matrix. The oscillation probability for one \neu flavour $\nu_\alpha$ into another one
$\nu_\beta$ is given by
\be
P (\nu_\alpha \rightarrow \nu_\beta) = \sint \times \sin^2 (1.27 \frac{L \times \Delta m^2 /eV^2}{E/MeV}) m 
\ee
with \delm = $m^2_j - m^2_i$ as the difference of two mass eigenstates $m_{i,j}$, $L$ the distance
from the neutrino source to the detector and $E$ the \neu energy.

\subsection{The LSND-evidence}
\begin{figure}
\begin{tabular}{cc}
\psfig{file=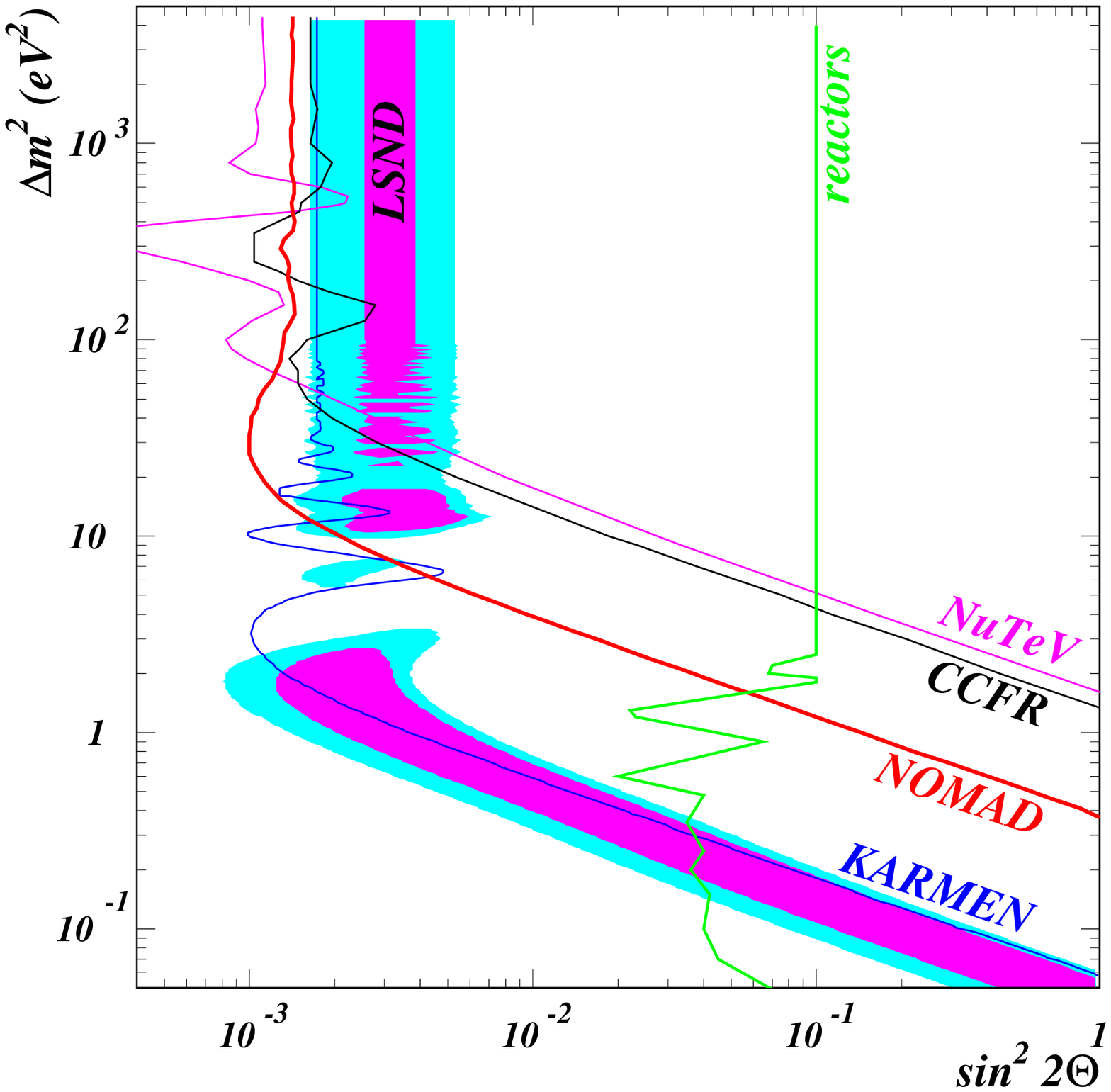,width=5cm} &
\psfig{file=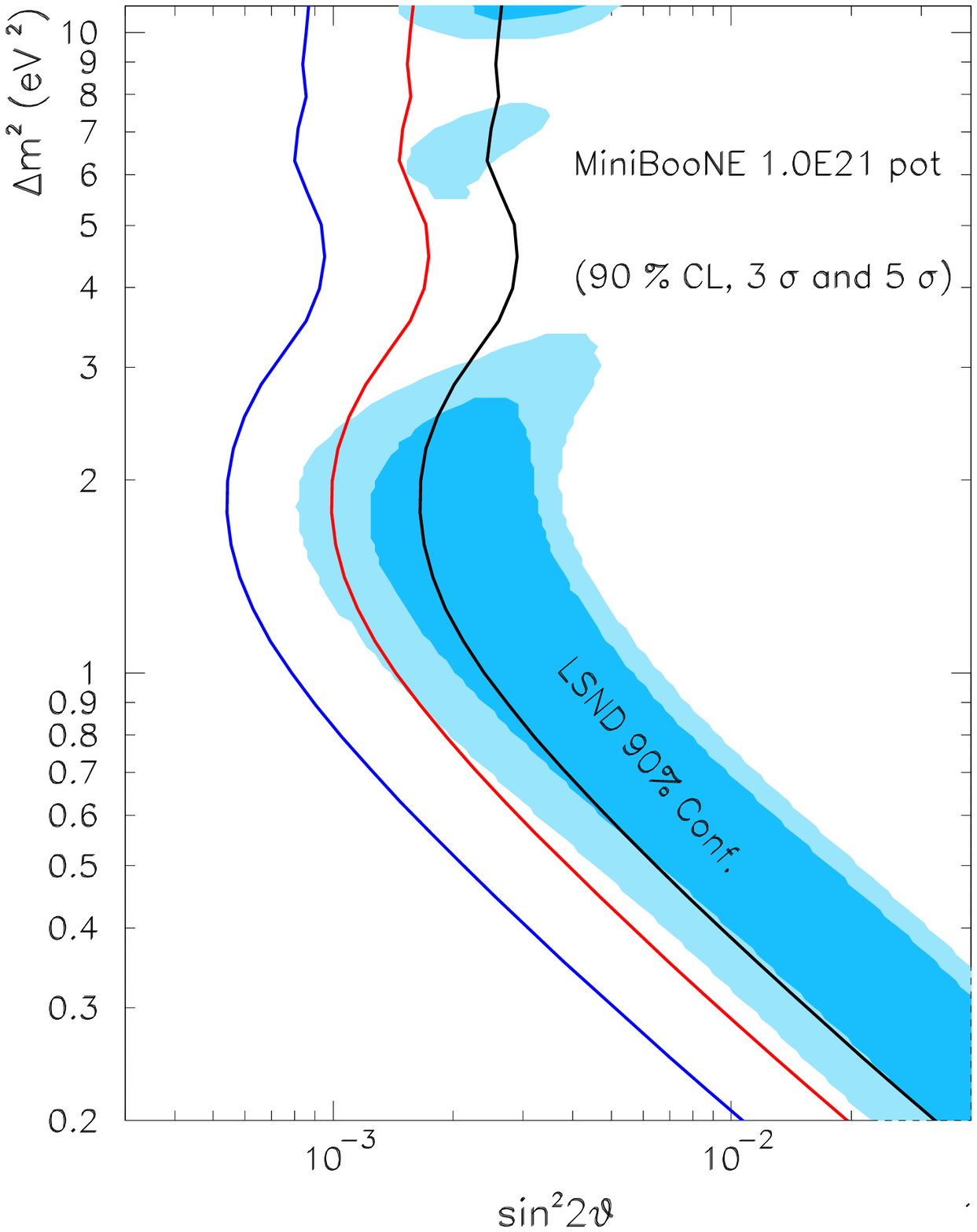,width=5cm}
\end{tabular}
\vspace*{8pt}
\caption{\label{lsnd} Left: \sint - \delm plot showing the parameter regions describing the LSND observation
(purple, blue). Exclusion curves from various other \exps are shown as well, with the region on the right side
excluded (from \protect \cite{ast03}). Right: In case of a non-observation of an effect in MiniBooNE the regions
on the right side of the curves can be excluded with the shown significance.}
\end{figure}

The \lsnd experiment at LANL was a 167 t mineral oil based
liquid
scintillation detector using
scintillation and Cerenkov light for detection. It consisted of an
approximately
cylindrical tank
8.3 m long and 5.7 m in diameter. LSND took data from 1993 - 1998.
For the ''decay at rest'' analysis in the channel 
\bnmu $\ra$ \bnel , the signal reaction was 
\be
\bnel + p \ra e^+ + n
\ee
As experimental signature a positron within the energy range 20 MeV
$<E_e<$ 60 MeV together
with a time and spatial correlated delayed 2.2 MeV photon from
p(n,$\gamma$)D is required. After background subtraction indeed an excess 
of $87.9 \pm 22.4 \pm 6.0$ events was observed \cite{agu01}. 
Interpreting those events as oscillation signal it would correspond to a transition
probability of $P ( \bnmu \ra \bnel) = 2.64 \pm 0.67 \pm 0.45 \times 10^{-3}$.\\
With rather similiar parameters the KARMEN experiment was operated at
Rutherford Appleton Laboratory from 1990 to 2001, finding no evidence for an oscillation 
signal \cite{arm02} (see figure \ref{lsnd}) .
To which extent both \exps are in agreement or not
is a severe statistical problem. A combined
analysis on both data sets has been performed \cite{chu02}. Two regions remain where
both \exps are compatible with a positive effect, one at \delm $\approx 7 eV^2$ and one with \delm $< 1 eV^2$.\\
With the \delm region known, it is possible to perform a new experiment to test this evidence.
The experiment MiniBooNE at Fermilab is exactly doing that.
The neutrino beam is produced by the Fermilab Booster, sending a 
high intensity pulsed proton beam of 8 GeV on a Be-target. The positively
charged secondaries, mostly pions, are focused by a magnetic horn and
brought into a decay tunnel. This results in an almost pure \nmu beam (\nel
contamination less than 0.3 \%). The detector itself is installed about 500m away from the  
end of the decay tunnel. It consists of 800 t of pure mineral oil, contained
in a 12.2 m diameter spherical tank. A support structure carries about 1550
phototubes for detection of Cerenkov and scintillation light. To explore the LSND evidence to
a level of 5$\sigma$ about 10$^{21}$ protons on target are required  (see figure \ref{lsnd}). 
More than 30 \% of the
data have been obtained and by 2005 first results can be expected. 

\subsection{Zenith angle dependence of atmospheric neutrinos and K2K}
\begin{figure}
\centerline{\psfig{file=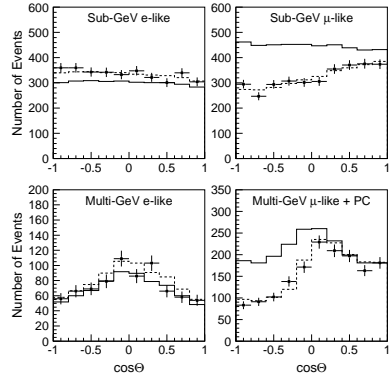,width=5cm}}
\vspace*{8pt}
\caption{\label{skzenit} Zenith angle distribution of electrons (left) and muons (right), both divided into low
and high energy samples. Clearly visibile is the deviation from Monte Carlo expectations (solid line) and data
points in the muon sample, especially for those coming from below ($\cos \theta < 0$. The dotted curve corresponds
to a fit including \neu \oscs (from \protect \cite{skatmos}).}
\end{figure}

\begin{figure}
\begin{tabular}{cc}
\psfig{file=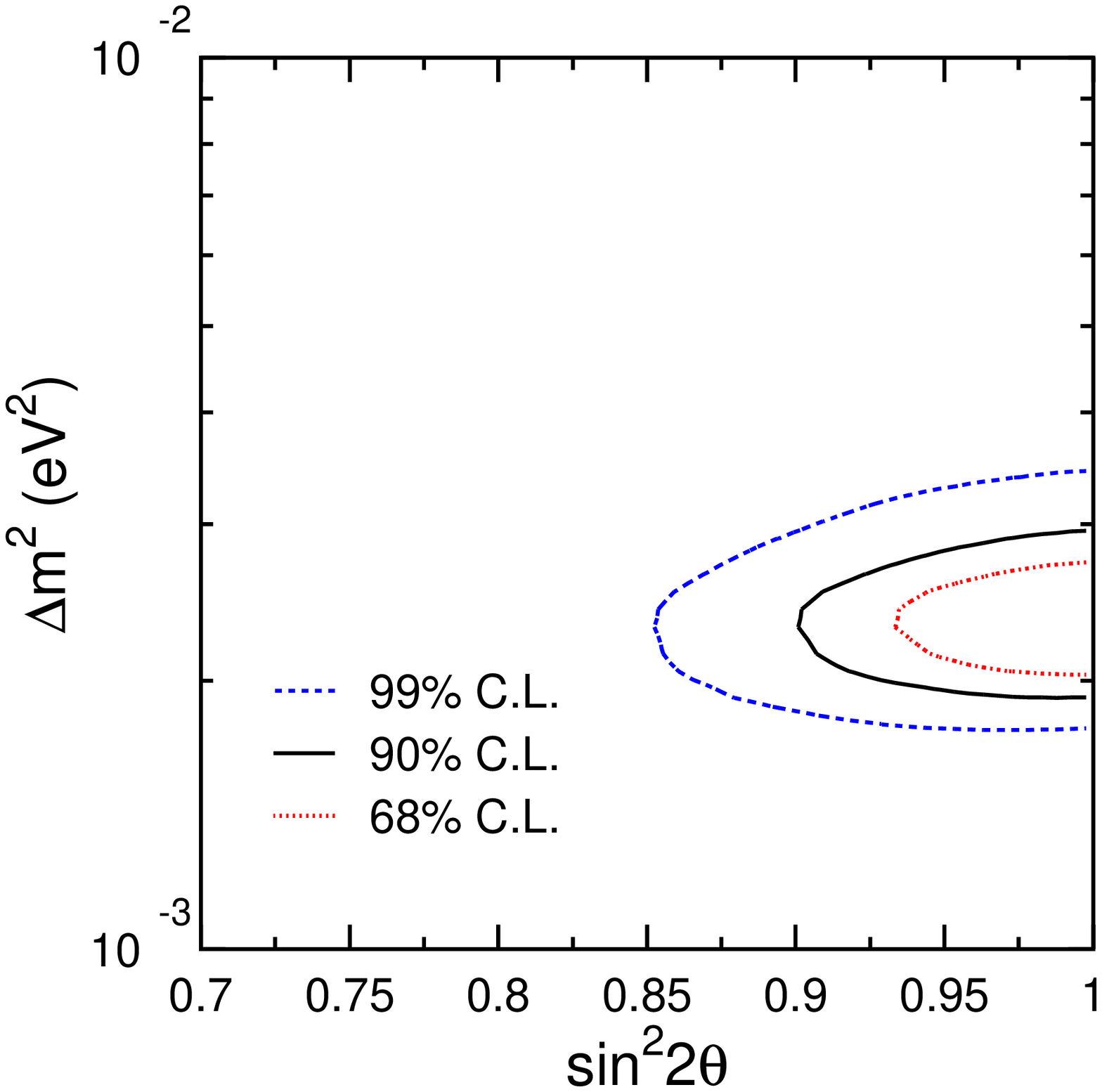,width=5cm} &
\psfig{file=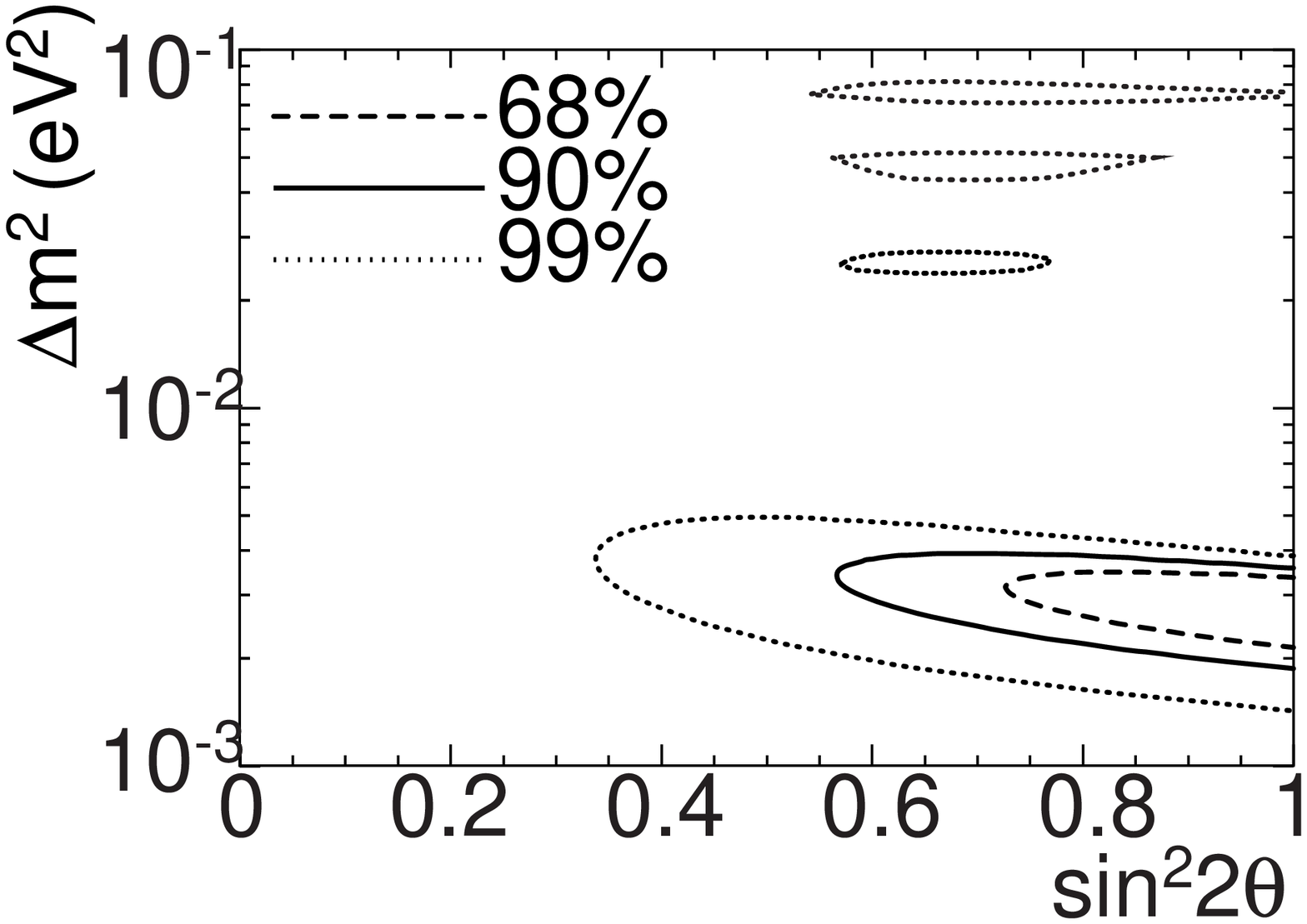,width=5cm}
\end{tabular}
\vspace*{8pt}
\caption{\label{atmosevi} Left: Contours obtained by the recent L/E analysis in Super-Kamiokande (from \protect
\cite{skle}). Right: 
Countours as obtained by K2K. Both are in reasonable agreement pointing towards and \delm between 2-3 $\times
10^{-3} eV^2$ and maximal mixing (from \protect
\cite{k2k}).}
\end{figure}

For more than a decade it is known that the ratio of electron/muon like events observed from 
atmospheric \nel and \nmu neutrinos does not agree with Monte Carlo expectation. A much deeper 
understanding has been obtained with the advent of Super-Kamiokande, which is able to perform
a measurement of the zenith angle distribution of both flavours separately (figure \ref{skzenit}) .
From that it can be concluded that
the reason for the deviation in the ratio is due to a lack of muons, or more precisely, the number
of upward going muons is reduced, an effect to be explained by neutrino oscillations including the 
\nmu. An involvement of \nel could be excluded by the CHOOZ and Palo Verde reactor experiments.
The parameters determined \cite{skatmos} are in agreement with maximal mixing and a \delm of $1.3-3 \times 10^{-3}$ eV$^2$.
Recently,  Super-Kamiokande has published a high resolution L/E analysis \cite{skle}, showing a better sensitivity to
the involved \delm. The outcome is a range of parameters as $1.9 \times 10^{-3} eV^2 < \delm < 3.0 \times 10^{-3} eV^2$
and \sint $>$ 0.90 with 90 \% CL and a best fit value of \delm = $2.4 \times 10^{-3} eV^2$ and \sint = 1.02 
(figure \ref{atmosevi}).\\
Independently the result is confirmed by the K2K experiment shooting a neutrino beam from KEK to Super-Kamiokande
with a baseline of 235 km. The current analysis \cite{k2k} is based on $8.9 \times 10^{19}$ protons on target and shows
a clear deficit in muon neutrinos (57 observed with 84.8 expected) suggesting a parameter region as shown in
figure \ref{atmosevi}. 
\\figure
Further long baseline experiments will be online soon, the next one is MINOS. This 5.4 kt magnetised iron spectrometer
is located in the Soudan mine in Minnesota using a neutrino beam from Fermilab. The baseline is 732 km 
and a low energy beam profile has been chosen first to have a good sensitivity for a disappearance search.
The detector is already operational and first beam is expected by end of 2004.
The European program, using a neutrino beam from CERN to the Gran Sasso Underground Laboratory in Italy 
is focussed on an optimized \ntau appearance search which implies a higher beam energy. Two experiments,
ICARUS and OPERA, are currently installed for the search. ICARUS will be a 3 kt LAr TPC working like an 
electronic bubble chamber and \ntau detection relies on the different distributions of kinematic variables 
in \nmu and \ntau charged current reactions \cite{rub01}. The basic building blocks of OPERA are sandwich sheets 
of lead and emulsions combined to form 8.3 kg bricks. In total more than 200000 bricks will be installed \cite{gul00}.
The excellent spatial resolution of emulsions allows to search for kinks within tracks, a characteristic 
feature of \ntau interactions. The data taking is foreseen to start in 2006 and within 5 years both
\exps should collect about a dozen $\tau$ candidates.
 
\subsection{Solar and reactor neutrinos}
Big progress has been achieved in the field of solar neutrinos. The problem of missing solar neutrinos 
has been solved by the
Sudbury Neutrino Observatory (SNO) measuring in their flavour-blind neutral current reaction on the deuteron
the expect flux from solar models \cite{sno}. Combined with all the other solar neutrino observations from \sk, Homestake and the
gallium experiments SAGE and GALLEX/GNO the parameter range could be pinned down to the large mixing angle solution,
showing that matter oscillations are responsible for the deficit in solar \nel. The current best fit value combining
all solar data is at \delm = $6.46 \times 10^{-5} eV^2$ and $\tan^2 \theta = 0.4$.

\begin{table}[h]
\tbl{Current results of solar neutrino experiments. Radiochemical results are given in SNU, fluxes from
water Cerenkov detectors in units of $10^6 cm^{-2}s^{-1}$.}
{\begin{tabular}{@{}cccc@{}} \toprule
Experiment & Target & Result & Prediction \\ \colrule
Homestake \hphantom{00} & \hphantom{0}$^{37}$Cl & \hphantom{0}2.56$\pm$ 0.23 & 7.6 $\pm$ 1.2\\
GALLEX/GNO \hphantom{00} & \hphantom{0}$^{71}$Ga & \hphantom{0}69.3$\pm$ 5.5 & 127 $\pm$ 10\\
SAGE \hphantom{00} & \hphantom{0}$^{71}$Ga & \hphantom{0}66.9$^{+5.3}_{-5.0}$ & 127 $\pm$ 10\\
Super-K \hphantom{00} & \hphantom{0}H$_2$O & \hphantom{0}2.35$\pm$ 0.10 & 5.1 $\pm$ 0.2\\
SNO & \hphantom{0}D$_2$O & 5.21 $\pm$ 0.47\hphantom{0}& 5.1 $\pm$ 0.2 \\ \botrule
\end{tabular}}
\end{table}

\begin{figure}
\begin{tabular}{cc}
\psfig{file=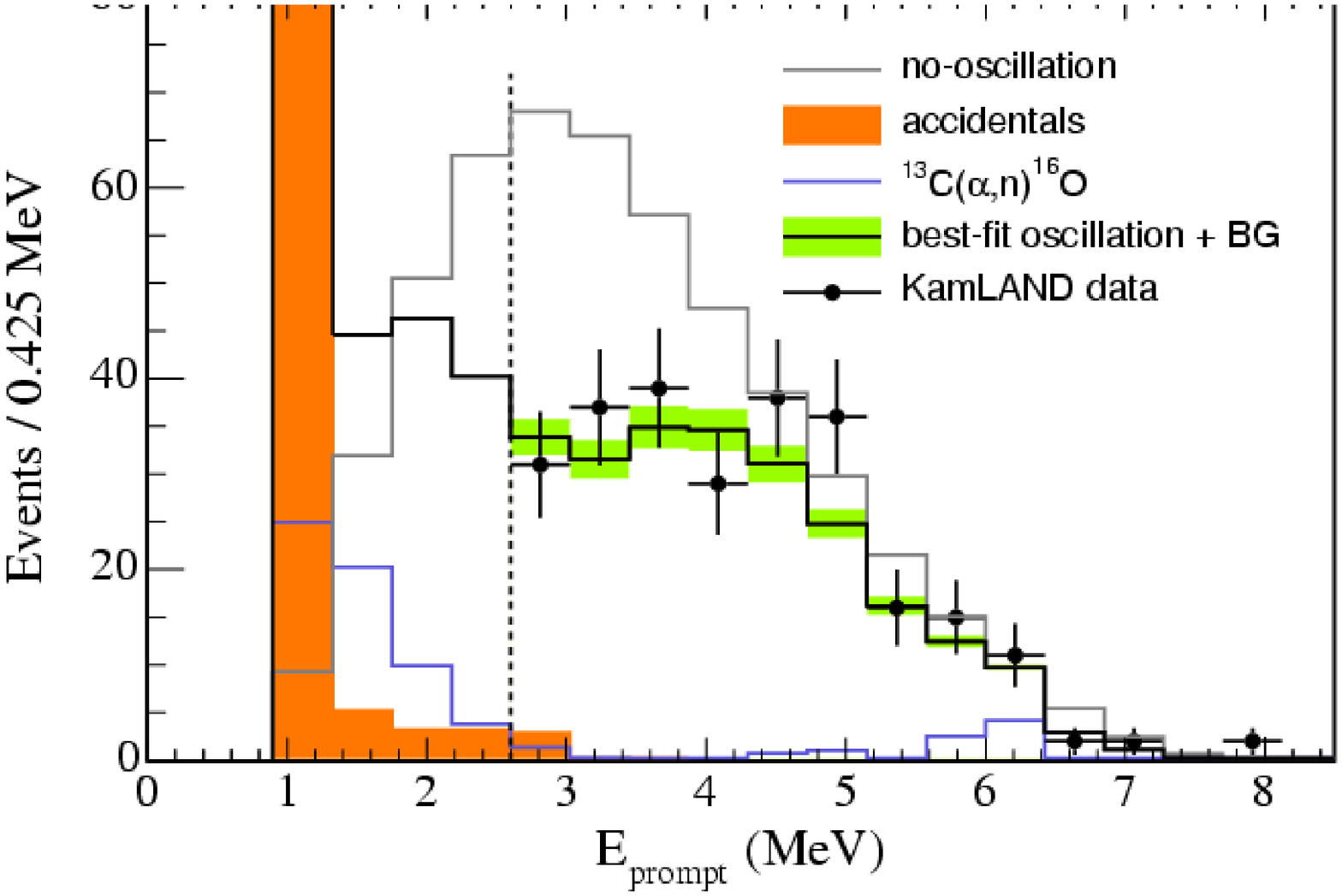,width=5cm} &
\psfig{file=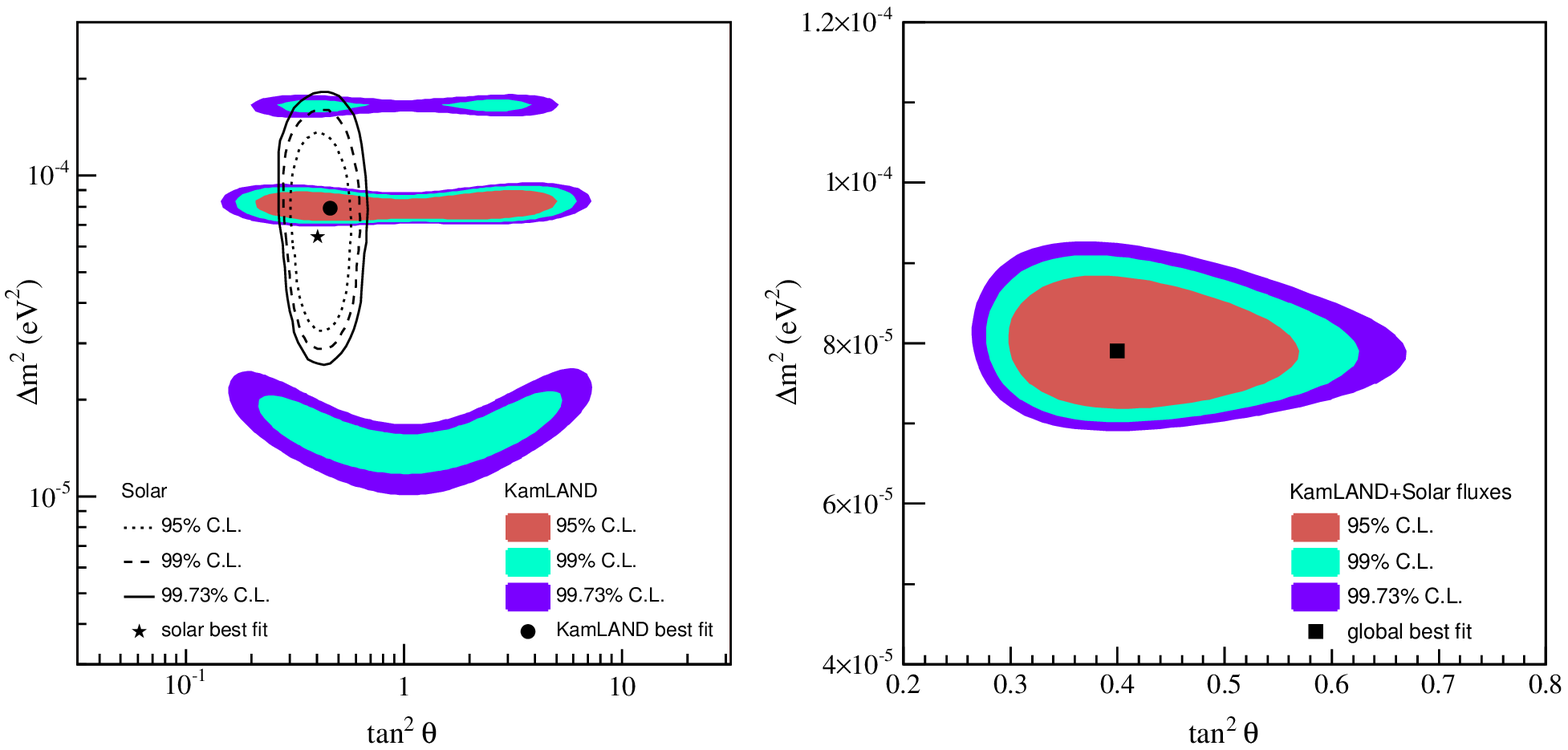,width=6cm}
\end{tabular}
\vspace*{8pt}
\caption{\label{solar} Left: Prompt energy spectrum as observed by KamLAND. A clear spectral distortion can be seen.
Right: Combined KamLAND and solar data fits (from \protect \cite{kamland}).}
\end{figure}

Completely independent information is coming from KamLAND, a long baseline experiment using nuclear power plants. 
The KamLAND experiment is a 1000 t Liquid Scintillator located at the former position of the Kamiokande detector in Japan.
After 515 days of data taking (766 ton $\times$ yr exposure) they see a clear spectral distortion and deficit of events
\cite{kamland}. 
The spectral distortions are very sensitive to \delm resulting in \delm = $7.9^{+0.6}_{-0.5}\times 10^{-5} eV^2$ 
hence the parameter space in combination with solar neutrinos in a global fit
could be further reduced to $\tan^2 \theta = 0.40 ^{+0.10}_{-0.07}$ (figure \ref{solar}).

To summarize, three evidences for \neu \oscs exist:
\begin{itemlist}
 \item The LSND-evidence, $10^{-3} < \sint <  10^{-1}$, $0.1 eV^2 < \delm < 6 eV^2$ , \nmu - \nel 
 \item The atmospheric zenith angle dependence \sint = 1.00, \delm = $2.4 \times 10^{-3} eV^2$, \nmu - $\nu_X$
 \item Solar and reactor neutrinos, \sint $\approx$ 0.81, \delm = $7.9 \times 10^{-5} eV^2$, \nel - $\nu_X$
\end{itemlist}

However, it is obvious that if all of them are correct, more neutrinos 
are needed, because of the unitarity of the mixing matrix, which only allows to form two independent \delm.
In combination with the LEP bound of three light neutrinos, those possible new neutrinos cannot participate in the standard
electroweak interactions and hence are called sterile.

\section{The determination of the elements of the full 3x3 mixing matrix}
The discussion of atmospheric and solar neutrinos in a two flavour scenario is justified by the fact that the
mixing angle $\sin^2 \theta_{13} < 0.12$ (90\% CL) at \delm $\approx 3 \times 10^{-3} eV^2$
as measured by the reactor experiments CHOOZ and Palo Verde. 
However, the knowledge of its
precise value and especially if it is non zero is extremely important, because only in this case it would allow the
possibility to search for CP-violation in the lepton sector. In the full three flavour mixing scenario the
mixing matrix U (called PMNS-matrix) is given by
\be
\label{pmns}
U_{PMNS} = \left( \begin{array}{ccc}
c_{12} c_{13} & s_{12} c_{13} & s_{13} e^{- i \delta}\\
- s_{12} c_{23} - c_{12}s_{23}s_{13} e^{i \delta}& c_{12}c_{23} -
s_{12}s_{23}s_{13}e^{i \delta} &
s_{23}c_{13}\\
s_{12}s_{23}- c_{12}s_{23}s_{13} e^{i \delta} & -c_{12}s_{23} - s_{12}c_{23}s_{13}e^{i
\delta} &  c_{23}c_{13}
\end{array} \right)
\ee
where $s_{ij} = \sin \theta_{ij}, c_{ij} = \cos \theta_{ij}\: (i,j=1,2,3)$.
In addition to the normally occuring CP-phase, there can be two more CP-violating phases $\alpha_1, \alpha_2$ 
associated with a possible \majo character,
which do not show up in oscillations but can have an effect in neutrinoless double beta decay.
The new matrix would be $U = U_{PMNS} \times $ diag $(1,e^{i \alpha_1},e^{i \alpha_2})$. As can be seen from eq.~\ref{pmns}
the CP-phase is always showing up in combination with $\sin \theta_{13}$ , which will ultimately determine the sensitivity
for CP-violation searches.\\
The precise oscillation probabilities in the three flavor scenario including matter 
effects are quite complex (see e.g. \cite{lin02,zub04}). One major result is the 8-fold degeneracy in parameters describing a specific oscillation
probablility, namely the degeneracy within the pairs $\delta - \sin 2 \theta_{13}, \delta - sign \Delta m^2_{13}$ and 
$\theta_{23} - (\pi /2 - \theta_{23})$.\\
The first step towards a search for a CP-violation will be a determination of \ted . Currently two strategies are 
followed, either using off-axis accelerator beams or performing a very precise new reactor experiment. A degeneracy
of neutrino parameters in off-axis beams makes a measurement at reactors desirable to disentangle the various parameters and 
break their degeneracy.\\
A determination of \ted{} at reactor search is coming from the survival probability
\be
P(\bnel \ra \bnel) = 1 - \sin^2 \theta_{13}  \sin^2 ( 1.27 \frac{\Delta m^2_{13}L}{E})
\ee
Taking the current oscillation evidences a baseline of about 1-2 km would show the maximum sensitivity. The precision
required including especially systematic effects, is pushing towards a two detector concept. A compilation of discussed
options and sites is shown in tab.~2.
\begin{table}[h]
\label{reactors}
\tbl{Proposed sites and parameters for considered reactor experiments to measure \ted.}
{\begin{tabular}{@{}ccccc@{}} \toprule
Site & Power & Baseline & Shielding & Sensitivity \\
& (GW$_{thermal}$) & (Near/Far) m & (Near/Far) mwe  & (90 \% CL) \\ \colrule
Krasnoyarsk, Russia\hphantom{00} & \hphantom{0}1.6 & \hphantom{0}115/1000 & 600/600 & 0.03 \\
Kashiwazaki, Japan\hphantom{00} & \hphantom{0}24 & \hphantom{0}300/1300 & 150/250 & 0.02 \\
Double Chooz, France\hphantom{0} & 8.4 & 150/1050\hphantom{0} & 30/300 & 0.03\hphantom{0} \\
Diablo Canyon, CA\hphantom{0} & 6.7 & 400/1700\hphantom{0} & 50/700 & 0.01\hphantom{0} \\
Angra, Brazil\hphantom{0} & 5.9 & 500/1350\hphantom{0} & 50/500 & 0.02\hphantom{0} \\
Braidwood, France\hphantom{0} & 7.2 & 200/1700\hphantom{0} & 450/450 & 0.01\hphantom{0} \\
Daya Bay, China & 11.5 & 250/2100\hphantom{0} & 250/1100 & 0.01\\ \botrule
\end{tabular}}
\end{table}

In addition to nuclear power plant searches, there is the option to use accelerator \neu beams off-axis. The important
point to notice here is the fact that due to the pion decay kinematics by going slightly off-axis the obtained neutrino
energy is basically independent from the original pion energy. Both are related via $E_{\nu} = 0.43 E_\pi / (1+ \gamma^2 \theta^2)$
with $\theta$ as the off-axis angle. Thus, one can obtain a narrow band beam at the expense of intensity. The \neu beam energies 
discussed are around 1 GeV and below. Here another important point enters, namely the precise knowledge of cross sections.
In this regime quasi-elastic scattering dominates with contributions from resonance production, coherent particle production and
diffractive interactions. A new proposal to accurately measure those cross-sections MINERvA at Fermilab. In addition, to get a good understanding
of the beam also the pion, kaon production in the target has to be known precisely. Experiments like HARP, NA49, MIPP have already
obtained data or will do so in the near future. Two proposals exist for off-axis beams, Nova and T2K, with the latter being approved.
While Nova plans to use the NuMI beam at Fermilab, T2K will be using the new accelerator facility in Tokai (Japan) to shoot a beam
towards Super-Kamiokande. In a second step this can be extended to a higher beam energy and a larger detector (Hyper-Kamiokande).
In addition to these superbeams, i.e. conventional neutrino beams with a high intensity, two completly new beam concepts
are envisaged as well. The first one is called beta beams. The idea is to accelerate $\beta$-unstable isotopes \cite{zuc02} 
to energies of a few 100 MeV using ion accelerators like ISOLDE at CERN. This would give a clearly defined beam of \nel or \bnel. 
Among the favoured isotopes discussed are $^6$He in case of a \bnel beam and $^{18}$Ne in case of a \nel beam. The second
one is a muon storage ring (''neutrino factory''). The two main advantages are the precisely known neutrino beam composition and
the high intensity (about 10$^{21}$ muons/year should be filled in the storage ring). Even if many technical challenges have
to be solved, it offers a unique source for future
accelerator based neutrino physics. First experimental steps towards
realisation are the HARP experiment at CERN, which determines the target for optimal 
production of secondaries, the study of muon scattering (MUSCAT experiment) 
and muon cooling (MICE experiment).

\section{Absolute \neu mass measurements}
\begin{figure}
\begin{tabular}{cc}
\psfig{file=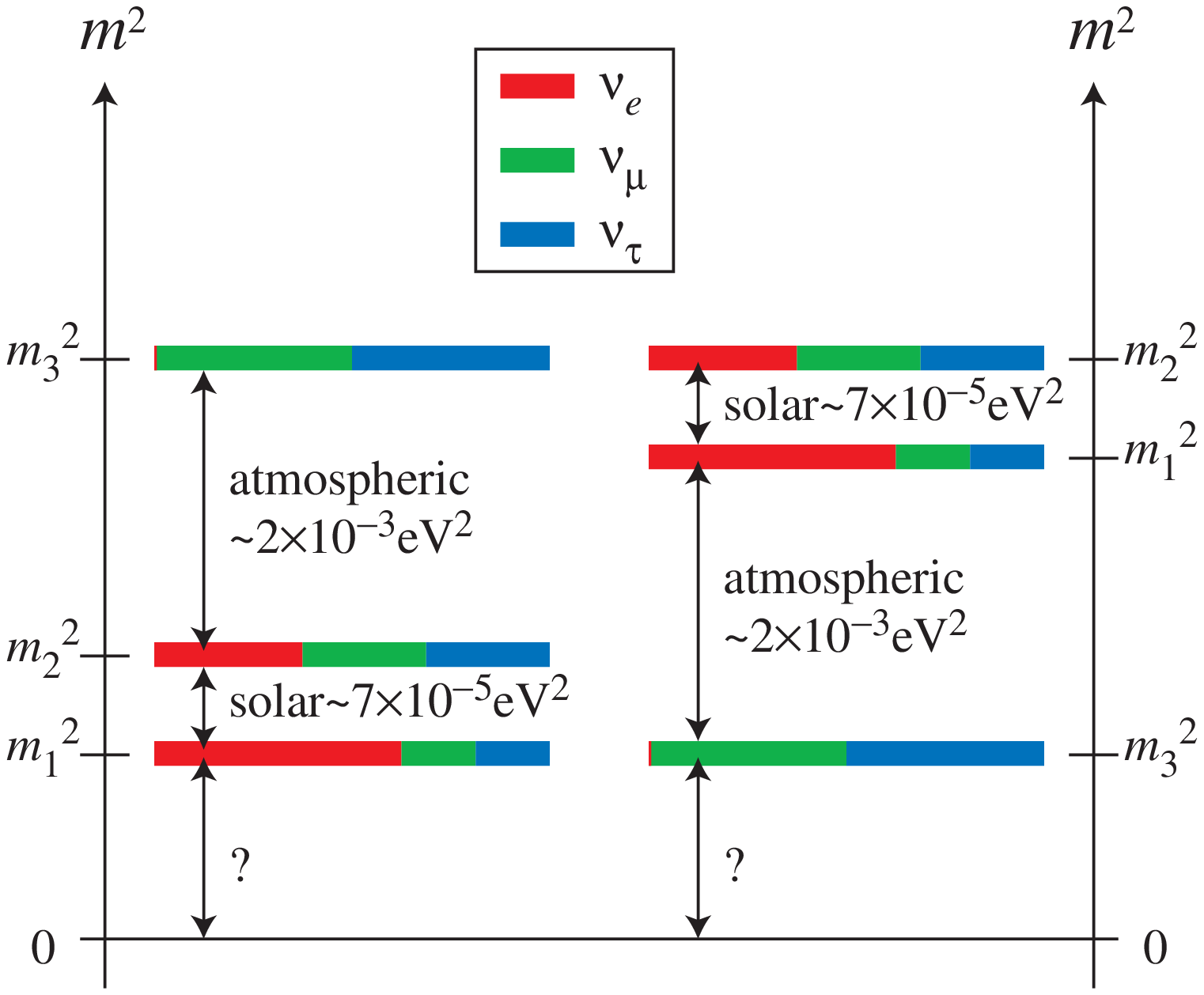,width=5cm} &
\psfig{file=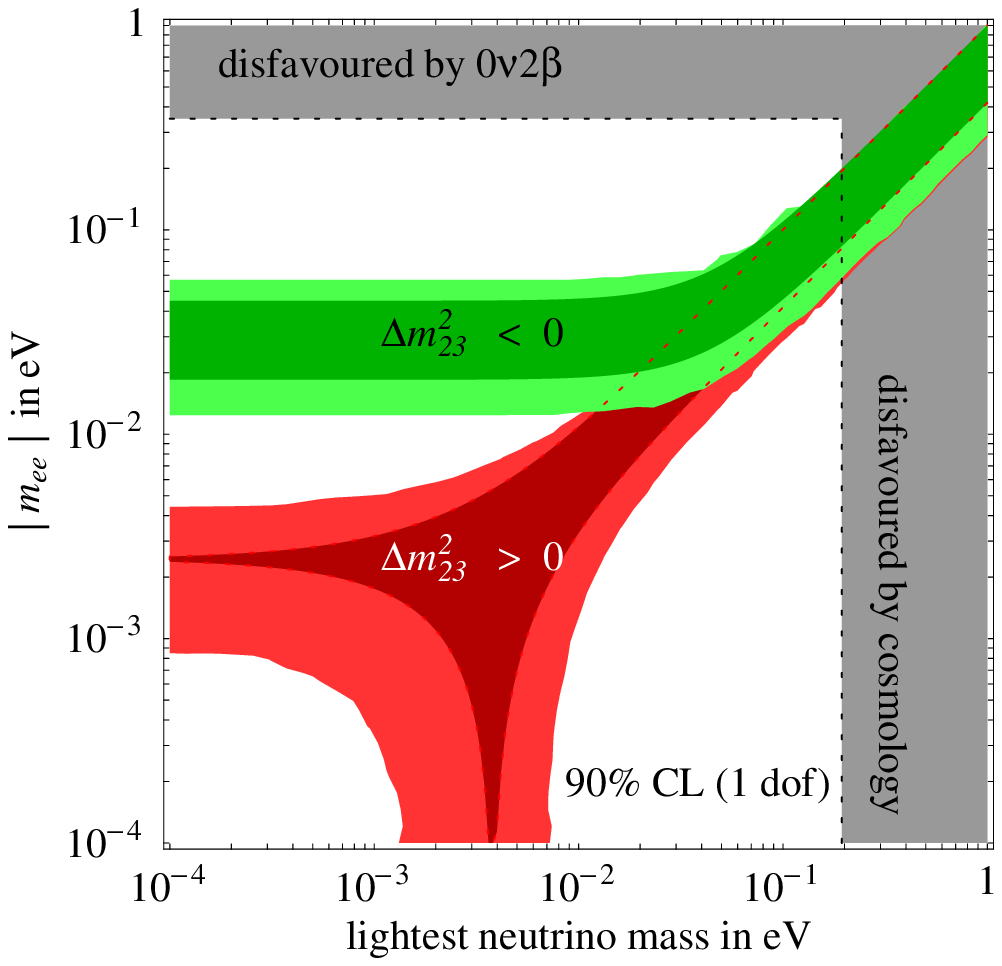,width=5cm}
\end{tabular}
\vspace*{8pt}
\caption{\label{masses} Left: Possible configurations of neutrino mass states as suggested by oscillations. Currently
a normal (left) and an inverted (right) hierarchy cannot be distinguished.  The flavour composition is shown as well.
Right: The effective \neu mass as a function the lightest neutrino mass state. As can be seen hierachical structure only
occur for mass well below 100 meV. If neutrino masses are above they are almost degenerated (from \protect \cite{fer02}).}
\end{figure}

Neutrino oscillations are no absolute mass measurements and thus allow for various
configurations of the mass eigenstates and mass models. Taking the small \delm involved,
for an absolute neutrinos mass above about 0.1 eV the \neu mass states will be almost degenerated.
However as function of the lightest mass eigenstate m$_1$ the two hierarchical models can be distinguished
for lower masses with the help of double beta decay (figure \ref{masses}). Currently three types of absolute mass 
determinations are explored, 
which have parts in common
but on the other hand also differences.

\subsection{Beta decay}
The precise investigation of the endpoint of the electron energy spectrum in tritium beta decay is the classical way to
search for a non-vanishing rest mass of the neutrino. Within the last decade due to new spectrometer developments 
two groups from Mainz and Troitzk were able to deduce an upper limit on the neutrino mass of $m_\nu<$2.2 eV
(95 \% CL). The actual measured quantity in the presence of mixing is
\be
m^2_{\nu_e} = \sum_i \mid U_{ei} \mid^2 m^2_{\nu_i}
\ee
A next generation of spectrometer, scaled in size to be sensitive to 0.2 eV, is KATRIN \cite{kat01}, currently under
installation in Germany.
It will start data taking in 2008. Two alternative ideas are the search using the beta emitter $^{187}$Re in compounds 
as cryogenic bolometers. The advantage is the very low Q-value of $^{187}$Re of only about 2.5 keV. In addition,
if a newly observed line in $^{115}$In is due to a beta-decay into an excited state, here the endpoint energy would be only about 
2 eV.

\subsection{Neutrinoless double beta decay}
\begin{figure}
\begin{tabular}{cc}
\psfig{file=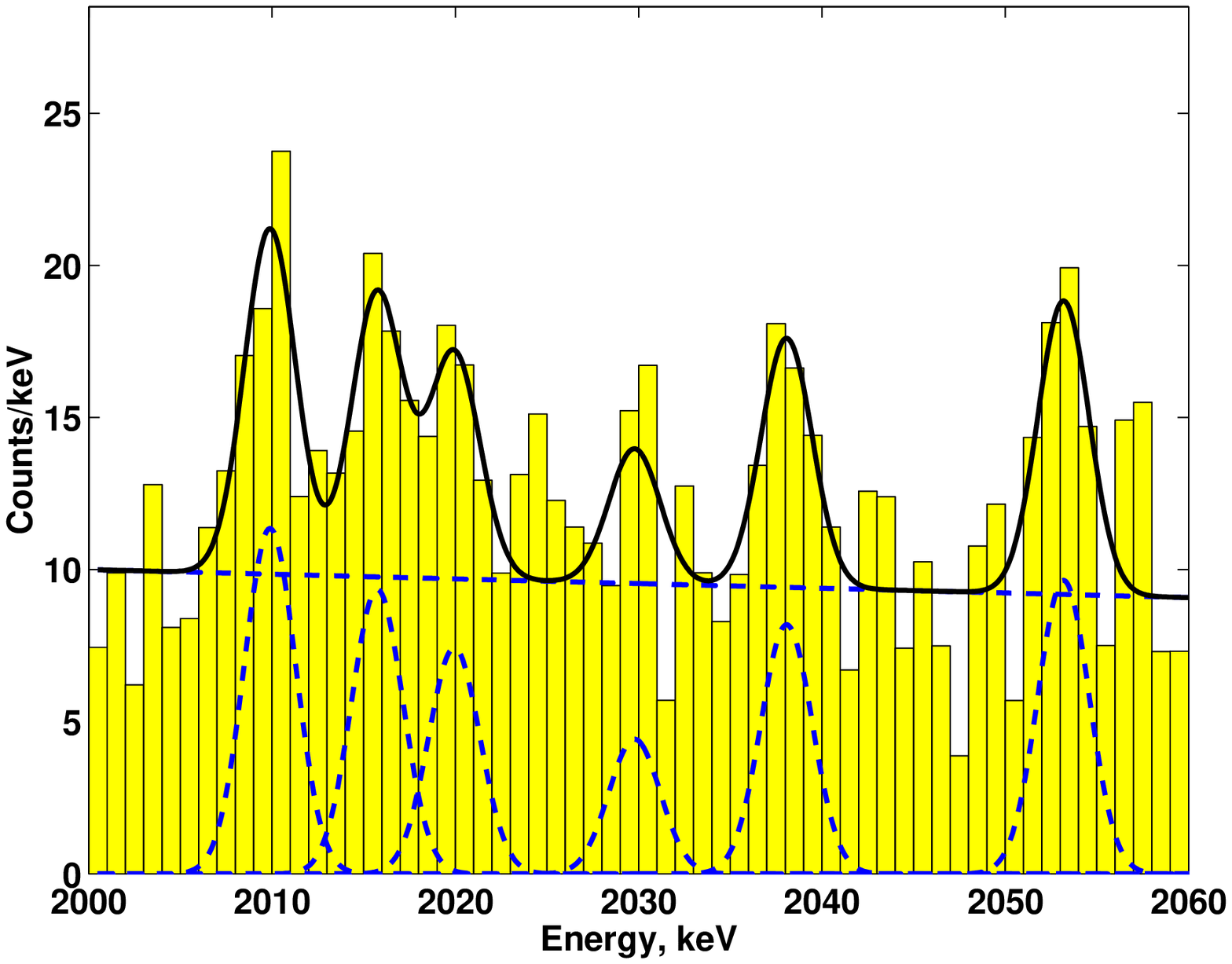,width=5cm} &
\psfig{file=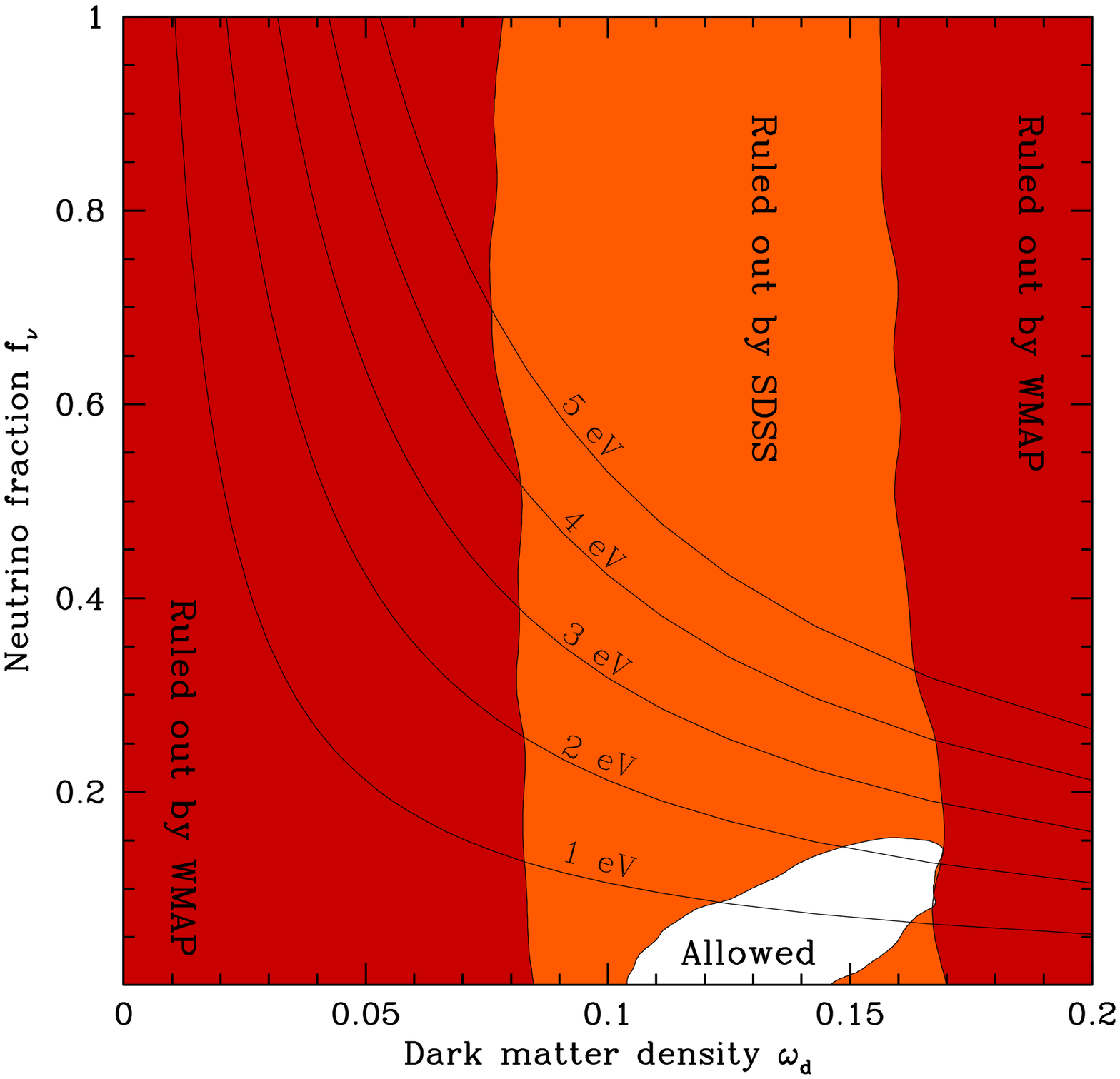,width=5cm} 
\end{tabular}
\vspace*{8pt}
\caption{\label{bbandlss}Left: Sum energy spectrum in the region around the double beta peak of $^{76}$Ge at
2039 keV as obtained by the Heidelberg-Moscow experiment(from \protect \cite{kkevidence}). 
Right: Neutrino mass fraction versus dark matter density
as an example of correlations among cosmological parameters. As can be seen the allowed range is up to about
the laboratory values (from \protect \cite{teg03}).}
\end{figure}

A different process related to neutrino masses is double beta decay. Of special importance is the neutrinoless decay mode
\be
(Z,A) \ra (Z+2,A) + 2 e^-  
\ee
which violates total lepton number by two units and requires massive \majo \neus and hence has sensitivity to the fundamental
character of \neus in contrast to beta decay. The experimental observable is a half-life which can be linked to the
\neu mass as 
\be
(T_{1/2})^{-1}
= G^{0\nu}(Q,Z) \mid M_{GT}^{0\nu} - M_F^{0\nu}\mid ^2 (\frac{\ema}{m_e})^2 
\ee
with $G^{0\nu}(Q,Z)$ as the well known phase space factors and $\mid M_{GT}^{0\nu} - M_F^{0\nu}\mid ^2$ as the involved nuclear matrix elements. The latter are a severe
source of uncertainty. 
The measured quantity is called effective \majo \neu mass and given by
\be
\ema = \mid \sum_i U_{ei}^2 m_i \mid =  \mid \sum_i \mid U_{ei} \mid^2 e^{2i\alpha_i} m_i \mid 
\ee 
The current situation is dominated by a hot debate of a claimed evidence (figure \ref{bbandlss}), 
as been observed with Ge-semiconductor detectors \cite{kkevidence}. The claimed half-life region of and the corresponding neutrino 
mass of would clearly show that \neus are almost degenerated. Currently two large-scale experiments are running, 
CUORICINO and NEMO-3. The first one is using 40 kg of Te$O_2$ as cryogenic bolometers and neutrino mass limits
in the range have been obtained \cite{arn03}. NEMO-3 is a TPC based detector using 10 kg of foils, dominantly $^{100}Mo$, for
the search and first results have been published recently \cite{arn04}.
Both experiments plan to upgrade their detectors towards larger masses. Of course there are further proposals and ideas
for future experiments.
Two proposals using enriched $^{76}$Ge are GERDA and MAJORANA. The first one is in a good situation to probe
the claimed evidence in a reasonable short time, by having the Heidelberg-Moscow and IGEX enriched Ge-detectors
at their hands. EXO, a He-filled TPC, focussing on the decay of $^{136}$Xe has 200 kg of enriched Xe and plans
to start measurement soon. COBRA \cite{zub01}, by using CdZnTe detectors the only other semiconductor approach besides Ge,
is operating several detectors at Gran Sasso Laboratory and has an enhanced sensitivity to double electron 
capture and double positron 
decays as well. A compilation of proposed experiments can be found in \cite{zub04}.

\subsection{Neutrino masses from cosmology}
During the last decade enormous progress has been made in observational cosmology and
the precision of current data allows to put some limits on neutrino masses. 
According to standard cosmology, in connection with the 3K cosmic microwave background (CMB)
there should exist a 1.96 K relic neutrino background. Taking the particle densities of 
both the well known relation for the density
\be
\Omega_\nu h^2 = \frac{m_{\nu, tot}}{94 eV}
\ee
can be obtained. Hence by measuring \omnu the sum of all three neutrino masses can be
obtained. Taking the upper limit from tritium beta decay and the oscillation results,
\neus can still contribute up to 15 \% of the total density. Cosmological bounds basically stem 
from large scale structure surveys. Neutrinos, being relativistic particles, at the beginning
of structure formation effectively washed out small scale perturbations. Hence, the net 
result is less small scale objects and thus a suppression in the power spectrum
given by \cite{hu98}
\be
\frac{\Delta P (k)}{P(k)} \approx -8 \frac{\Omega_\nu}{\Omega_m}
\ee
should be observed. However, all mass bounds obtained are depending on the cosmological model used. Currently
the standard lore is a $\Lambda$CDM model with adiabatic linear perturbations. In addition,
the other cosmological parameters have to be known, currently determined by CMB observations.
The reason is that there is a strong correlation of $\Omega_\nu$ with other cosmological 
parameters \cite{han02}. Depending on the assumptions and data used limits on the neutrino mass between
0.3-3 eV have been obtained \cite{lah04}, even a non-vanishing rest mass could be obtained showing the strong
dependence on the assumptions. A comprehensive recent analysis on CMD and SDSS data can be found in \cite{teg03}.
To sum it up, it is fair to say, that cosmological bounds achieved the same level of sensitivity
as laboratory experiments. In the future the comparison of all three areas, beta decay,
neutrinoless double beta decay and cosmological mass determinations will improve and it
will be very exciting to explore their consistency and gain further information on the
neutrino mass. 


\section*{Summary and conclusion}
Neutrino physics has made major progress over the last decade. A non-vanishing rest mass has been
established in oscillation experiments. The solar neutrino problem is solved in being due to 
matter oscillations, independently confirmed by nuclear reactor searches. However, the three 
evidences do not all fit together and if all are true, more neutrinos than those from the
Standard Model are needed. The next step will be to more precisely determine the elements of
the full 3x3 mixing matrix, ultimately trying to detect CP-violation in the leptonic sector. The first
step to do is a more precise determination of the angle \ted{} in nuclear reactor searches and
off-axis beams. \\
Neutrino oscillations do not determine the absolute mass scales. For that beta decay, neutrinoless
double beta decay and cosmological studies can be used. In all three areas major progress has been
achieved and can be expected in the next decade. 

\section*{Acknowledgments}
This work is supported by a Heisenberg Fellowship of the Deutsche
Forschungsgemeinschaft (DFG).


\begin{thebibliography}{0}


\bibitem{agu01} A. Aguilar et al, LSND collaboration, {\it Phys. Rev. D} 
{\bf 64}, 112007 (2001).

\bibitem{arm02} B. Armbruster et al, KARMEN collaboration,  {\it Phys. Rev. D} 
{\bf 65}, 112001 (2002).

\bibitem{chu02} E. D. Church et al, {\it Phys. Rev. D} 
{\bf 66}, 013001 (2002).

\bibitem{ast03} P. Astier et al, NOMAD collaboration {\it Phys. Lett. B} 
{\bf 570}, 19 (2003).

\bibitem{skatmos}M. Ishitsuku, Preprint hep-ex/040676

\bibitem{skle} Y. Ashie et al, Super-Kamiokande collaboration, {\it Phys. Rev. Lett.} 
{\bf 93}, 101801 (2004).

\bibitem{k2k} E. Aliu et al, K2K collaboration, Preprint hep-ex/0411038.

\bibitem{sno} S. N. Ahmed et al, SNO collaboration, {\it Phys. Rev. Lett.} 
{\bf 92}, 181301 (2004).

\bibitem{rub01} A. Rubbia, {\it Nucl. Phys. B (Procs. Suppl.)} 
{\bf 91}, 223 (2001).

\bibitem{gul00} M. Guler et al , OPERA Proposal, LNGS P25/2000, CERN SPSC 2000-028

\bibitem{kamland} T. Araki et al., KamLAND collaboration, Preprint hep-ex/040635

\bibitem{lin02} M. Lindner, Preprint hep-ph/0209083

\bibitem{zub04} K. Zuber, {\it Neutrino Physics} 
(IOP Publ., Bristol, 2004).

\bibitem{zuc02} P. Zucchelli,  {\it Phys. Lett. B} 
{\bf 532}, 166 (2002).

\bibitem{fer02} F. Feruglio, A. Strumia, F. Vissani,  {\it Nucl. Phys. B} 
{\bf 637}, 345 (2002), ibid. {\bf 659},359 (2003).

\bibitem{kat01} Proposal KATRIN experiment, Preprint hep-ex/0109033

\bibitem{kkevidence} H. V. Klapdor-Kleingrothaus et al,  {\it Phys. Lett. B} 
{\bf 586}, 198 (2004).

\bibitem{arn03} C. Arnaboldi et al, CUORICINO collaboration, Preprint hep-ex/0302006

\bibitem{arn04} R. Arnold et al, NEMO3 collaboration  {\it JETP Lett. } 
{\bf 80}, 429 (2004).

\bibitem{zub01} K. Zuber {\it Phys. Lett. B} 
{\bf 519}, 1 (2001).

\bibitem{hu98} W. Hu, D. Eisenstein and M. Tegmark, {\it Phys. Rev. Lett.}
{\bf 80}, 5255 (1998).

\bibitem{han02} S. Hannestad {\it Phys. Rev. D}
{\bf 66}, 125011 (2002).

\bibitem{lah04} O. Lahav, O. Elgaroy, Preprint astro-ph/0411092, astro-ph/0412075

\bibitem{teg03} M. Tegmark et al, {\it Phys. Rev. D}
{\bf 80}, 103501 (2003).


\end{thebibliography}
\end{document}